\documentclass[referee]{raa}           
\usepackage{graphicx,times}
\usepackage{natbib}
\usepackage{amssymb,amsmath}
\bibpunct{(}{)}{;}{a}{}{,}
\usepackage{longtable,lscape}
\usepackage[pagebackref=true]{hyperref}
\usepackage[paperheight=26cm,paperwidth=22cm,right=3cm,left=3cm,top=3cm,bottom=3cm]{geometry}%font a4
\begin{document}

   \title{A catalog of collected debris disks: properties, classifications and correlations between disks and stars/planets}
%\subtitle{properties; classifications; correlations between disks and stars/planets}
 \volnopage{ {\bf 20XX} Vol.\ {\bf X} No. {\bf XX}, 000--000}
   \setcounter{page}{1}

   \author{
   	Peng-cheng Cao\inst{1}, Qiong Liu\inst{1,2}, Neng-Hui Liao\inst{1}, Qian-cheng Yang \inst{1},  Dong Huang\inst{1}
   }
%% Here is an example of three authors come from different institutes.
%% For single author or all the authors from an institute, use "\inst{}" only

   \institute{ Department of Physics, Guizhou University, Guiyang 550025, China; {\it qliu1@gzu.edu.cn}; {\it nhliao@gzu.edu.cn}\\
%% Please give the E-mail address of the author, to whom future correspondence and
%% offprint requests will be sent.
        \and
             Institute of Astronomy, University of Cambridge, Madingley Road, Cambridge CB3 0HA, UK\\
\vs \no
 %  {\small Received 20XX Month Day; accepted 20XX Month Day}
}
\abstract{
We have collected a catalog of 1095 debris disks with properties and classification (resolved, planet, gas) information. From the catalog, we defined a less biased sample with 612 objects and presented the distributions of their stellar and disk properties to search for correlations between disks and stars. We found debris disks were widely distributed from B to M-type stars while planets were mostly found around solar-type stars, gases were easier to detect around early-type stars and resolved disks were mostly distributed from A to G-type stars. The fractional luminosity dropped off with stellar age and planets were mostly found around old stars while gas-detected disks were much younger. The dust temperature of both one-belt systems and cold components in two-belt systems increased with distance while decreasing with stellar age. In addition, we defined a less biased planet sample with 211 stars with debris disks but no planets and 35 stars with debris disks and planets and found the stars with debris disks and planets had higher metallicities than stars with debris disks but no planets. Among the 35 stars with debris disks and planets, we found the stars with disks and cool Jupiters were widely distributed with age from 10 Myr to 10 Gyr and metallicity from -1.56 to 0.28 while the other three groups tended to be old ($\textgreater$ 4Gyr) and metal-rich ($\textgreater$ -0.3). Besides, the eccentricities of cool Jupiters are distributed from 0 to 0.932 wider than the other three types of planets ($\textless$ 0.3).
}
   \authorrunning{P. C. Cao et al. }            %author_head in even pages
   \titlerunning{A catalog of collected debris disks}  % title_head in odd pages
   \maketitle

%________________________________________________ sections below
% 

\section{Introduction}           %% first-level sections will be auto-capitalized
\label{sect:intro}

Circumstellar disks are common products of star formation. At the early stage of star formation, there is a lot of primordial gas and dust left over which are called protoplanetary disks. These disks seem to dissipate after a few million years \citep{2005ApJ...620.1010R}. With the amount of this material decreasing, the disks evolved from protoplanetary disks to transitional disks to debris disks \citep{2008ApJ...676..509R}. Debris disks are dusty, gas-poor disks around main-sequence stars and contain little or no primordial material. In 1983, the first extrasolar debris disk was detected around Vega by  Infrared Astronomy Satellite (IRAS) \citep{1984ApJ...278L..23A}. It is known that many main-sequence stars show infrared and/or submillimeter excess emission which is thought to be re-radiation from the debris dust grains \citep{2008ApJ...674.1086T}. These grains are replenished by collisions between the remnants of the planet formation process \citep{2008ARA&A..46..339W, 2009ApJ...700L..73K, 2010RAA....10..383K}. 	

Over the past few decades, dozens of published papers focus on searching for stars with infrared or submillimeter excess emission which are based on the following telescopes: IRAS, Wide-field Infrared Survey Explorer (WISE), Infrared Space Observatory (ISO), AKARI, Herschel, Spitzer, and Atacama Large Millimeter/sub-millimeter Array (ALMA) \citep{2007ApJ...660.1556R, 2016ApJS..225...15C, 2001ApJ...555..932S, 2014AJ....148....3L, 2016A&A...593A..51M, 2014ApJS..211...25C, 2022AJ....164..100S}. So far, thousands of debris disk candidates around main-sequence stars have been detected which display diverse properties. 

The study of debris disks contains many aspects. Early works focus on searching for new debris disk candidates. 
\citet{2007ApJ...660.1556R} confirmed 146 stars showed excessive emission at 60 $\mu$m through IRAS, \citet{2014ApJS..211...25C} identified 499 debris disks through Spitzer/IRS (Infrared Spectrograph), and \citet{2016A&A...593A..51M} analysed a sample of 177 stars through Herschel and got 33 targets with infrared excesses. 

With more and more disks being detected, later works intend to search for correlations between disks and their host stars. From the spectral energy distribution (SED) \citep{2011ApJ...730L..29M, 2013ApJ...775...55B, 2014ApJS..211...25C}, the dust emission can often be fitted by one or two temperatures from tens to hundreds of kelvins which represents that the disk usually have one or two discrete rings \citep{2013ApJ...775...55B}. Warm dust is the most powerful tool for studying the inner regions of habitable planets which provides a method to detect the formation of planets \citep{2009ApJ...701.2019L}, and cold dust can be used to study the formation or migration of planets in the outer region of planetary systems. \citet{2011ApJ...730L..29M} analysed the SEDs of 69 debris disks, including 50 A-type stars and 19 Solar-type stars. Comparing the properties of warm dust around these two sample stars, they found both have similar warm disk temperatures (190 K). \citet{2013ApJ...775...55B} found 225 targets with significant excess: 100 with a single cold component, 51 with a single warm component, and 74 with both warm and cold components. They found a positive trend between cold debris disk temperature and stellar temperature. 

At the same time, there are many works dedicated to determining the correlation between planets and debris disks.
\citet{2007MNRAS.380.1737W} predicted that debris disks detected around planet-hosting stars should be more infrared luminous than those around stars without planets. 
Since then, several efforts have been devoted to studying this correlation. \citet{2009ApJ...705.1226B} collected a sample of 146 known radial velocity (RV) planet-hosting stars and 165 unknown planet-hosting stars while \citet{2009ApJ...700L..73K}  collected a sample of 150 planet-hosting stars and 118 stars without planets with Multiband Imaging Photometer (MIPS) observations, they both found that stars with planets do tend to have brighter debris disks than those stars without planets, although the difference was not statistically significant. 
And later, \citet{2012MNRAS.424.1206W} found some evidence for a correlation between debris disks and low-mass planets through analysis of a sample of the nearest 60 G-type stars. While \citet{2015ApJ...801..143M} did not found evidence that debris disks were more common or more dusty around stars harboring high-mass or low-mass planets compared to a control sample without identified planets. 
More recently, \citet{2020MNRAS.495.1943Y} came to a similar conclusion by studying 201 known RV planet-hosting stars and 294 unknown planet-hosting stars which showed that there was no significant correlation between RV planet presence and debris disk properties. As we know, both the debris disks and planets are believed to have been formed in protoplanetary disks. It can be expected that the properties of these two components should be related to each other in some way \citep{2018haex.bookE.165K}. 

Besides, there are many articles devoted to studying the composition and structure by detecting the gas component and imaging disks. \citet{2017ApJ...849..123M} detected CO gas in disks of three A-type stars using ALMA telescope, \citet{2012A&A...546L...8R} found OI gas in a debris disk through Herschel/PACS (Photoconductor Array Camera and Spectrometer), and \citet{2017MNRAS.469..521K, 2019MNRAS.489.3670K} used the secondary origin model to explain the gas origin of most disks. And the number of debris disks with gas detection is still increasing. The study of gas can help us better understand the evolution of debris disks and the formation of planets, as well as the study of debris disk structure. 
Disk structure can be seen clearly from images of resolved disks such as the two-belt systems \citep{2012ApJ...750L..21B, 2014ApJ...780...97M, 2015ESS.....312013R, 2022A&A...664A.139D, 2022ApJ...933L...1M}.
\citet{2019ApJ...877L..32M} found the gap structure in HD15115 through direct imaging and thought it might be caused by planet sculpting. The same event also occurred in HD92945 \citep{2019MNRAS.484.1257M}.

The study of debris systems helps us to better understand the formation and evolution of planetesimal belts and planetary systems \citep{2004ApJ...603..738Z, 2011ApJS..193....4M, 2014prpl.conf..521M}. With larger debris disk samples, we can better understand the characteristics of debris disks. Therefore, in this paper, we aim to build a catalog by collecting samples from published literature and search for their correlations to host stars and planets. We will describe the collected debris disk samples in Section 2 and study the properties’ distribution and correlations between debris disks and host stars in Section 3. Then in Section 4, we will discuss the classification of debris disks and study the correlation between disks and planets; Finally, we give the conclusion in Section 5.
 
\section{Samples collected from published literature}
\label{sect:Sample}

To establish a large debris disk catalog, we collect over 100 articles for stars claimed to have debris disks based on pivotal investigations including IRAS, WISE, ISO, AKARI, Spitzer, and Herschel. 
In the following subsections, we show some of the samples discovered by these telescopes.

\subsection{The IRAS discovered debris disks samples}
Since the discovery of the first debris disk around Vega, more and more debris disks were identified by infrared excess using IRAS data.

\citet{1998ApJ...497..330M} cross-correlated the Michigan Catalog of Two-dimensional Spectral Types for the HD Stars with the IRAS Faint Source Survey Catalog and identified a sample of 108 debris disk candidates. Whilst, there had no further information about the disk including the fractional luminosity.

\citet{2007ApJ...660.1556R} cross-correlated the Hipparcos stars with the IRAS catalogs and identified 146 stars within 120 pc of Earth that show excess emission at 60$\mu$m. We will refer to this sample as Rhee07 sample, hereafter. There is a wealth of parametric information about stars and disks including fractional luminosity, dust temperature, and location. Moreover, their investigation focuses on the mass, dimensions, and evolution of dusty debris disks which helps to understand the evolution of the planetary system. While there are 45 stars with only an IRAS detection at 60$\mu$m and with only upper limits at 25 and 100$\mu$m,  the dust properties of these stars still cannot be determined appropriately.

As a consequence, using more sensitivity data of WISE, 
\citet{2021RAA....21...60L} (hereafter Paper I) refitted the excessive flux densities of Rhee07 sample. Paper I got the dust temperature and fractional luminosities of these 45 stars which cannot be determined due to the limitations with the IRAS database (have IRAS detection at 60$\mu$m only), meanwhile, revised the dust properties of the remaining stars and found that the dust temperatures were overestimated in the high-temperature band. Therefore, We collected the Rhee07 sample classified as IRAS discovered disks but used the dust properties of Paper I.

\subsection{The WISE discovered debris disks samples}
As a more sensitive telescope, WISE plays its supplement role in the field of the debris disk.
\citet{2016ApJS..225...15C}  cross-correlated the Tycho-2 with AllWISE catalogs and identified 74 new sources of excess in their “Prime” catalog.  Notably, this study produced almost a 20$\%$ increase in debris disks as well as a handful of very dusty disks ($L_{IR}/L_{*} > 0.01$).

\subsection{The ISO discovered debris disks samples}
Compared to IRAS, ISO has expanded our capability to search for debris disks with an order of magnitude increase in sensitivity and a factor of 2 improvement in resolution. 
Using ISO, Spangler et al. (2001) detected 33 stars with debris disks of which 20 were new discoveries and showed the fractional luminosity dropped off with age which was consistent with a power-law index of 2.

\subsection{The AKARI discovered debris disks samples}
With a spatial resolution better than IRAS, AKARI/FIS will significantly reduce the false contamination in comparison with IRAS.
\citet{2014AJ....148....3L} (hereafter Paper II) cross-correlated the Hipparcos main-sequence star catalog with the AKARI/FIS (Far-Infrared Surveyor) catalog and identified 75 stars with debris disks of which 32 stars are new. Though the sensitivity of AKARI/FIS is similar to IRAS and both samples (this sample in Paper II and Rhee's sample in \citet{2007ApJ...660.1556R} ) are based on Hipparcos catalog, there are only 27 stars in common. And due to the shallow sensitivity limit of AKARI/FIS, this study can only recover the brightest debris disks. 

\citet{2017A&A...601A..72I} cross-correlated the Hipparcos and Tycho-2 catalog with AKARI/IRC (Infrared Camera) catalog and identified 53 debris disk candidates including 8 new detections. This study detected some faint warm debris disks around nearby stars. At least 9 objects showed large excess emissions for their ages, which would be challenging the conventional steady-state collisional cascade model.

\subsection{The Spitzer discovered debris disks samples}

Unlike the previous all-sky surveys IRAS and AKARI, Spitzer covers much smaller areas of the sky at mid- and far-infrared bands but possesses much better spatial resolutions and sensitivities. 
So the samples based on Spitzer are not cross-correlated from a stellar catalog such as Hipparcos and Tycho-2. 

\citet{2006ApJ...653..675S} measured 160 early-type main-sequence stars using Spitzer/MIPS of which 137 stars have 70$\mu$m observations and identified 44 stars with 70$\mu$m excess. This work focused on the evolution of debris disks and found that older stars tend to have lower fractional dust luminosity than younger ones. 

\citet{2008ApJ...674.1086T} observed 193 solar-type stars using Spitzer/MIPS and identified 31 stars with 70$\mu$m excess. Combined with previously published results,  this work found the excess rate for sun-like stars is 16.4$\%$ at 70$\mu$m.

\citet{2010ApJ...712.1421S} observed 71 stars using Spitzer/MIPS of which 37 solar-type stars are in the Pleiades and identified 23 stars with 24$\mu$m excess.  Notably, this study of the Pleiades probes the debris disk characteristics in a large sample of solar-type stars of well-determined age. 

\citet{2013ApJ...775...55B} accumulated a sample of 546 stars with Spitzer/IRS observations and identified 225 targets with significant excess. This work found the trend between the temperature of the inner edges of the cold debris disks component and the stellar type of their host stars is inconsistent with theories that cold debris disk location is strictly temperature-dependent.  Then they ruled out the dominance of ice lines in sculpting the outer regions of planetary systems and found no evidence that delayed stirring causes the trend. 

\citet{2014ApJS..211...25C} compiled the largest catalog with 499 debris disks with Spitzer/IRS excess. We will refer to this sample as Chen14 sample, hereafter. They found that the SEDs of about 66$\%$ targets can be fitted well by a two-temperature model with warm (100–500 K) and cold (50–150 K) dust components. Notably, this work gives the trends that younger stars generally have disks with larger fractional luminosities and higher dust temperatures, and higher-mass stars have disks with higher dust temperatures.  These trends will be tested in the following sections.

\subsection{The Herschel discovered debris disks samples}

\citet{2013A&A...555A..11E} presented the observational results of the Herschel Open Time Key Project DUNES.
A total of 31 out of the 133 DUNES targets show excess above the photospheric predictions including 10 new discoveries.
Notably, this work gives a significant fraction of the resolved disks and some more faint Far-Infrared excesses disks.
Moreover, they presented a weak trend of the correlation of disk sizes and the anti-correlation of disk temperatures with the stellar age. 

\citet{2016A&A...593A..51M} analysed a sample of 177 stars with 20 pc from two Open Time Key Programmes, DUNES, and DEBRIS, and got 33 targets with infrared excesses.
This work gave the incidence of debris disks around FGK stars in the solar neighbourhood per spectral type for different subsamples.
The results presented that the incidence of debris disks was similar for active (young) and inactive (old) stars.

\subsection{The collected catalog and sample bias}
After collecting all debris disks from the above samples and individual sources from other literature, we got a catalog of 1095 debris disks (hereafter total sample) which are listed in Table \ref{table:1}.  

\clearpage
\begin{center}
{\scriptsize
	\renewcommand{\footnoterule}{}
\begin{longtable}{llllllclclccll}
\caption{\label{table:1}The stellar and disk properties of our catalog sources.}\\
					\hline
					Name      & M/H  & D   & Spt & Age &   \multicolumn{2}{c}{One Temperature} &  \multicolumn{4}{c}{Two Temperature}  &$L_{IR}/L_{*}$ &Ref   &Telescope     \\
					\cline{6-9} 
					  \cline{9-11}					
					&        &&&
					& $T_{d}$  & $L_{IR}/L_{*}$   & $T_{d1}$   & $L_{IR,1}/L_{*}$ & $T_{d2}$   & $L_{IR,2}/L_{*}$  & &&\\
										\cline{1-14}
				         &                 &(pc)  &    &   (Myr) & (K)   &    & (K)  &   & (K)  &  &      &       &  \\	
					\cline{1-14}
					(1)&(2)&(3)&(4)&(5)&(6)&(7)&(8)&(9)&(10)&(11)&(12)&(13)&(14)	\\
					\hline
				HD 105                        & -0.45       & 38.83    & G0V   & 30         & 48  & 1.90E-04              & ... & ...                    & ... & ...                    & 1.90E-04 & 1         & Spitzer   \\   
				HD 166$^{r}$                        & 0.08        & 13.77    & G8    & 20         & ... & ...                   & 310 & 4.10E-05               & 72  & 4.10E-05               & 8.20E-05 & 1         & Spitzer   \\   
				HD 203                        & -0.21       & 39.74    & F3V   & 12         & ... & ...                   & 499 & 4.20E-06               & 127 & 1.00E-04               & 1.04E-04 & 1         & Spitzer   \\   
				HD 377$^{r}$                       & -0.23       & 38.40    & G2V   & 170        & ... & ...                   & 240 & 5.80E-05               & 57  & 2.00E-04               & 2.58E-04 & 1         & Spitzer   \\   
				HD 432                        & ...         & 16.78    & F2III & 1000       & 140 & 1.10E-04              & ... & ...                    & ... & ...                    & 1.10E-04 & 1         & Spitzer   \\   
				HD 870                        & -0.08       & 20.65    & K0V   & 5980       & ... & ...                   & 499 & 2.10E-05               & 54  & 1.50E-05               & 3.60E-05 & 1         & Spitzer   \\   
				HD 1051A                      & -0.72       & 131.15   & A7III & 600        & 36  & 2.67E-04              & ... & ...                    & ... & ...                    & 2.67E-04 & 7,61      & IRAS      \\   
				HD 1237A$^{p}$                      & -0.47       & 17.55    & G8V   & 11900      & 300 & 1.63E-04              & ... & ...                    & ... & ...                    & 1.63E-04 & 11,19     & WISE      \\   
				HD 1404                       & ...         & 41.32    & A2V   & 469        & ... & ...                   & 249 & 1.10E-05               & 115 & 6.40E-06               & 1.74E-05 & 1         & Spitzer   \\   
				HD 1461$^{p}$                       & 0.02        & 23.40    & G3V   & 6880       & ... & ...                   & 499 & 3.20E-05               & 54  & 2.90E-05               & 6.10E-05 & 1         & Spitzer   \\   
				HD 1466$^{p}$                       & -0.43       & 42.82    & F8V   & 30         & ... & ...                   & 374 & 7.90E-05               & 97  & 5.20E-05               & 1.31E-04 & 1         & Spitzer   \\   
				HD 1562                       & -0.61       & 24.73    & G1V   & 6166       & 75  & 5.80E-05              & ... & ...                    & ... & ...                    & 5.80E-05 & 11,23     & WISE      \\   
				HD 1581                       & -0.35       & 8.61     & F9.5V & 3020       & 218 & 1.60E-05              & ... & ...                    & ... & ...                    & 1.60E-05 & 41        & Spitzer   \\   
				HD 1835                       & -0.02       & 21.33    & G2.5V & 600        & ... & ...                   & 329 & 4.10E-05               & 75  & 6.70E-06               & 4.77E-05 & 1         & Spitzer   \\   			
					\hline
	\end{longtable}
	Notes. Column 1 is the name of host stars with upper right marks for the classification of their debris disks: p = debris disks with planets, g = gas-detected debris disks and r = resolved debris disks.
Columns 2 is the star's metallicity, with data from Gaia DR3; Columns 3 and 4 are the stellar distance from Earth and spectral type, with data from SIMBAD. Column 5 is the stellar age collected from published literature listed in Column 13. 	
Columns 6 to 11 are the disk properties: dust temperature $T_{d}$ and fractional luminosity $L_{IR}/L_{*}$. 
Column 12 is the total fractional luminosity $L_{IR}/L_{*}$ . 
In column 13, we listed the reference of Columns 6 to 13:
(1)\citet{2014ApJS..211...25C}; (2)\citet{2018ApJ...869L..40S}; (3)\citet{2021ApJ...910...27M}; (4)\citet{2001ApJ...555..932S}; (5)\citet{2007ApJ...660.1556R}; (6)\citet{2010ApJ...712.1421S}; (7)\citet{2012ApJ...745..147R}; (8)\citet{2010ApJ...710L..26K}; (9)\citet{2006ApJ...644L.125S}; (10)\citet{2013ApJ...775...55B}; (11)\citet{2016ApJS..225...15C}; (12)\citet{2014AJ....148....3L}; (13)\citet{2009ApJ...700L..73K}; (14)\citet{2009ApJ...698.1068P}; (15)\citet{2016ApJ...826..123M}; (16)\citet{2013ApJ...773..179W}; (17)\citet{2005ApJ...620.1010R}; (18)\citet{1998ApJ...497..330M}; (19)\citet{2017A&A...601A..72I}; (20)\citet{2016A&A...593A..51M}; (21)\citet{2006ApJS..166..351C}; (22)\citet{2013A&A...555A.104A}; (23)\citet{2012A&A...541A..40M}; (24)\citet{2006ApJ...653..675S}; (25)\citet{2013ApJ...768...25G}; (26)\citet{1993prpl.conf.1253B}; (27)\citet{2013A&A...551A.134O}; (28)\citet{2016MNRAS.456..459D}; (29)\citet{2014ApJ...785...33S}; (30)\citet{2008ApJ...676..509R}; (31)\citet{2017ApJ...849..123M}; (32)\citet{2013A&A...556A.119B}; (33)\citet{1996MNRAS.279..915S}; (34)\citet{2006ApJ...644..525M}; (35)\citet{2013A&A...555A..11E}; (36)\citet{2009ApJ...705...89L}; (37)\citet{2016RMxAA..52..357G}; (38)\citet{2009ApJ...705.1226B}; (39)\citet{1987A&A...181...77C}; (40)\citet{2019ApJ...870...36M}; (41)\citet{2008ApJ...674.1086T}; (42)\citet{2007ApJ...658.1289T}; (43)\citet{2008A&A...487.1041A}; (44)\citet{2014MNRAS.445.2558T}; (45)\citet{2015RMxAA..51....3G}; (46)\citet{2008ApJ...677..630H}; (47)\citet{2011A&A...536L...4E}; (48)\citet{1986PASP...98..685S}; (49)\citet{2015ApJ...801..143M}; (50)\citet{2013ApJ...773...73J}; (51)\citet{2008ApJ...679..720C}; (52)\citet{2014A&A...565A..68R}; (53)\citet{2012ApJ...753..147D}; (54)\citet{2016ApJ...817L...2C}; (55)\citet{2017A&A...600A..62P}; (56)\citet{2006ApJ...653.1480W}; (57)\citet{2001A&A...365..545H}; (58)\citet{2019MNRAS.488.3588Y}; (59)\citet{2004ApJ...608..526L}; (60)\citet{2020ApJ...894..116L}; (61)\citet{2021RAA....21...60L};(62)\citet{2012PASP..124.1042M};(63)\citet{2015ApJ...798...87M};(64)\citet{2021ApJ...912..115H};(65)\citet{2018MNRAS.475.3046S}.
Column 14 is the telescope used to detect the debris disk.
(This table is available in its entirety in a machine-readable form in the online journal. A portion is shown here for guidance regarding its form and content.)
}
 \end{center}

%-------------------------------------------------------------
While the samples are collected from different telescopes, their infrared fluxes have observational biases because of the sensitivity difference. The sensitivity of each telescope is listed in Table \ref{table:2}, from which we can see IRAS and AKARI are less sensitive compared to Spitzer and Herschel. Therefore, we should carefully check the sources detected by IRAS or AKARI. In order to avoid or reduce the bias, we only study the sources with complete disk properties for these sources have more than one band of infrared data. Furthermore, in order to discuss the correlation between disks and their host stars, we need to know the stellar properties which are not complete for every source, especially the stellar age. As a result, we define a prime debris disk sample with complete stellar and disk properties from the catalog.

Among 1095 stars in the total sample, there are 612 stars (hereafter debris disk sample) with complete stellar and disk properties including stellar age, spectral type, distance, fractional luminosity and dust temperature. Details of these properties are discussed in the following section. The numbers of sources with complete/incomplete properties among different telescopes are listed in Table \ref{table:2}.  From the table, we can see the sources in the debris disk sample are mostly (516/612) detected by the most sensitive telescope Spitzer. For the 46 stars detected by IRAS, we find that most of them (39/46) are in our paper I whose SEDs have been refitted by combining the WISE data and the remaining 7 stars have been detected by Spitzer, WISE or James Clerk Maxwell Telescope (JCMT). For the 2 stars detected by AKARI, the SED fitting also combined the WISE data. Therefore, we only study the debris disk sample with 612 stars in the following sections instead of the total sample with 1095 stars. As a word of caution, the conclusion drawn from these 612 stars cannot apply to the total sample.		

%\clearpage
	\begin{longtable}{lcccccc}
		\caption{\label{table:2}The different telescopes used to detect debris disks in the catalog. }\\
		\hline
		Telescope & Band($\mu$m) & Sensitivity (mJy) &No. Complete & No. Incomplete & Sum   \\
		\hline
		IRAS & 60 & 600 & 46 & 113 &159 \\
		WISE & 22 & 6 &14 & 102&116\\
		ISO & 60 & 3.6 &0 & 21 &21 \\
		AKARI & 90 & 550 & 2 & 71 &73 \\
		Spitzer & 70 & 6 & 516&  162&678 \\
	        Herschel & 70 & 4.4& 34 & 14 &48 \\
		\hline
		Total & ... & ... & 612 & 483 &1095 \\
		\hline
	\end{longtable}
	Notes. Column 1 is the telescope, Columns 2 and 3 are the commonly used infrared band and corresponding sensitivity, Column 4 is the number of sources with complete properties, Column 5 is the number of sources without complete properties, Column 6 is the number of sources detected by the corresponding telescope.
%-------------------------------------------------------------

\section{Properties and correlations between debris disks and host stars}
\label{sect:results}
In the last section, we defined a less biased debris disk sample base on complete properties. Then in this section, we will introduce the origins of these properties in detail and then study the distributions of stellar and disk properties and correlations between debris disks and host stars. 

\subsection{Stellar properties}

Firstly, we obtain the stellar properties including stellar age, spectral type and distance from the Earth (D). 
The distance and spectral type information are obtained from SIMBAD \citep{2000A&AS..143....9W}. 
Stellar age is useful information to understand dust properties since debris disks dissipate with time \citep{2005ApJ...620.1010R} and are expected to evolve dynamically with time \citep{2004AJ....127..513K}. Therefore, we also collected age information from published literature.  In order to study the metallicity distribution in the following section, we cross-matched the debris disk samples with Gaia DR3 catalog \footnote{\it https://archives.esac.esa.int/gaia} and got 475 sources with metallicity information. We put all information of the above stellar properties in Table \ref{table:1}. 

\begin{figure}
	\centering
	\includegraphics[width=16cm,angle=0]{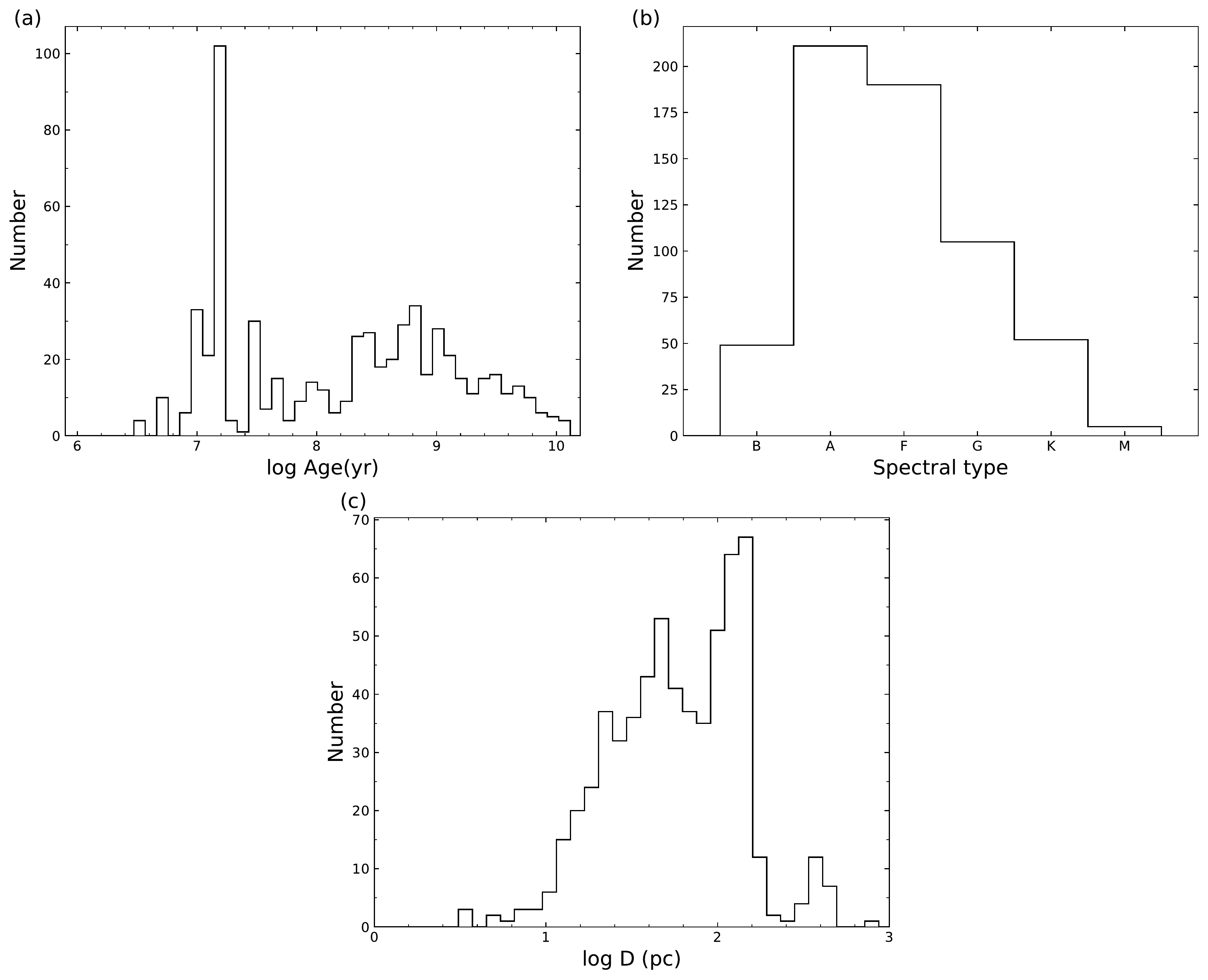}
	\caption{(a) Age histogram showing the number of targets observed as a function of age. (b) Spectral type histogram showing the number of targets observed as a function of spectral type. (c) Distance histogram showing the number of targets observed as a function of distance.}
	\label{figure1}
\end{figure}

Then, with the above stellar properties, we can study their distributions including stellar age, spectral type and distance. From Figure \ref{figure1} (a), we can see that about half of the stars are less than 100 Myr while the other half including a quarter is 100 Myr to 1 Gyr old, and the remaining quarter is older than 1 Gyr.
From Figure \ref{figure1} (b), we can see there are more solar-type stars (347 stars) than early-type stars (260 stars). Such spectral classification will be used in further discussion.
From Figure \ref{figure1} (c), the sample stars are mostly nearby, located within 200 pc of the Earth. 

\subsection{Disk properties}

Secondly, we collect disk properties including fractional luminosity $L_{IR}/L_{*}$ and dust temperature $T_{d}$ from published literature. For the source that appears in more than one paper, we carefully examine the disk properties and select the latest data.
According to the collected temperature information, we classify
the 612 stars in the debris disk sample into two groups: one-belt systems (281 sources are better fit using a one-temperature model) and two-belt systems (331 sources are better fit using a two-temperature model). 
Among the 612 stars, there are 480 sources in Chen14 sample. Note both one and two-belt systems in Chen14 sample have assumed minimum and maximum dust temperatures of 30 and 500 K.

\begin{figure}
	\centering
    \includegraphics[width=16cm,angle=0]{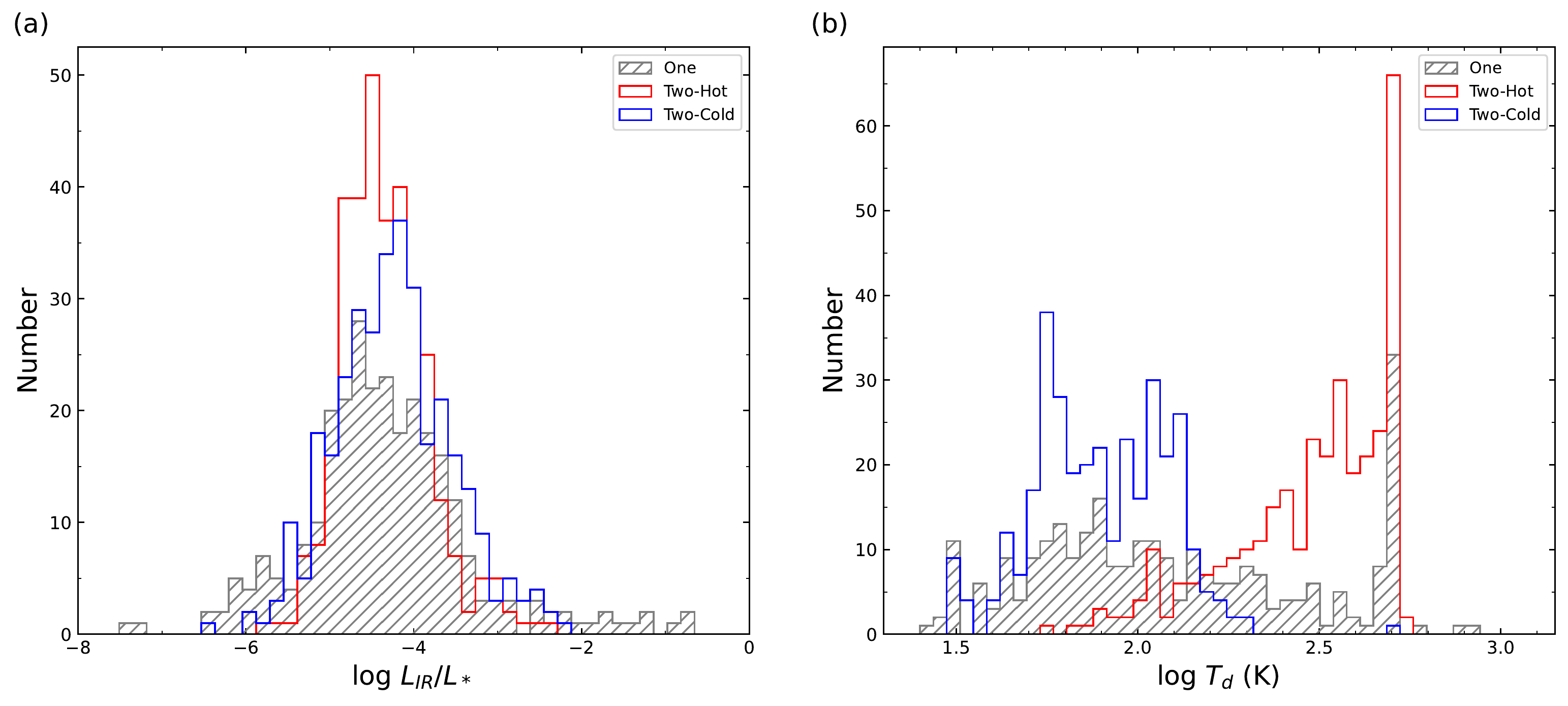}
	\caption{(a) Fractional luminosity histogram showing the number of targets observed as a function of fractional luminosity. (b) Dust temperature histogram showing the number of targets observed as a function of dust temperature. The grey histogram shows the distribution for one-belt systems; The red and blue histograms show the distributions of the warm and cold components in two-belt systems, respectively.}
	\label{figure2}
\end{figure}

The distributions of $L_{IR}/L_{*}$ and $T_{d}$ are shown in Figure \ref{figure2} (a) and (b), respectively. 
Comparing the distributions of one-belt systems and two-belt systems, the fractional luminosity has an obvious difference as Figure \ref{figure2} (a) shows. The one-belt systems are widely distributed from $10^{-7}-10^{-1}$ and two-belt systems are narrowly distributed from $10^{-6}-10^{-3}$, but there is no obvious difference between cold and warm components in two-belt systems.
From Figure \ref{figure2} (b), We find that the temperature of warm dust and cold dust in the two-belt systems mostly fall into the range of 100-500K and 30-160K, respectively. As a word of caution, these ranges are influenced by the minimum and maximum dust temperatures of 30 and 500 K in the Chen14 sample. The dust temperature of asteroid belt ($\sim$230K) and Kuiper belt ($\sim$40K) in our solar system falls into this range. While there is no obvious difference between one-belt and two-belt systems. 

\subsection{Correlations between debris disks and host stars}
Last but not least, we search for correlations between different properties of debris disks and host stars. In order to reduce the possible bias from different spectral types, we divide the debris disk sample with 612 objects into narrow spectral type ranges: early-type sample with 260 objects (88 one-belt systems and 172 two-belt systems) and solar-type sample with 347 objects (188 one-belt systems and 159 two-belt systems). Note we do not discuss the remaining 5 M stars which is a too-small sample.

\begin{figure}
	\centering
	\includegraphics[width=16cm,angle=0]{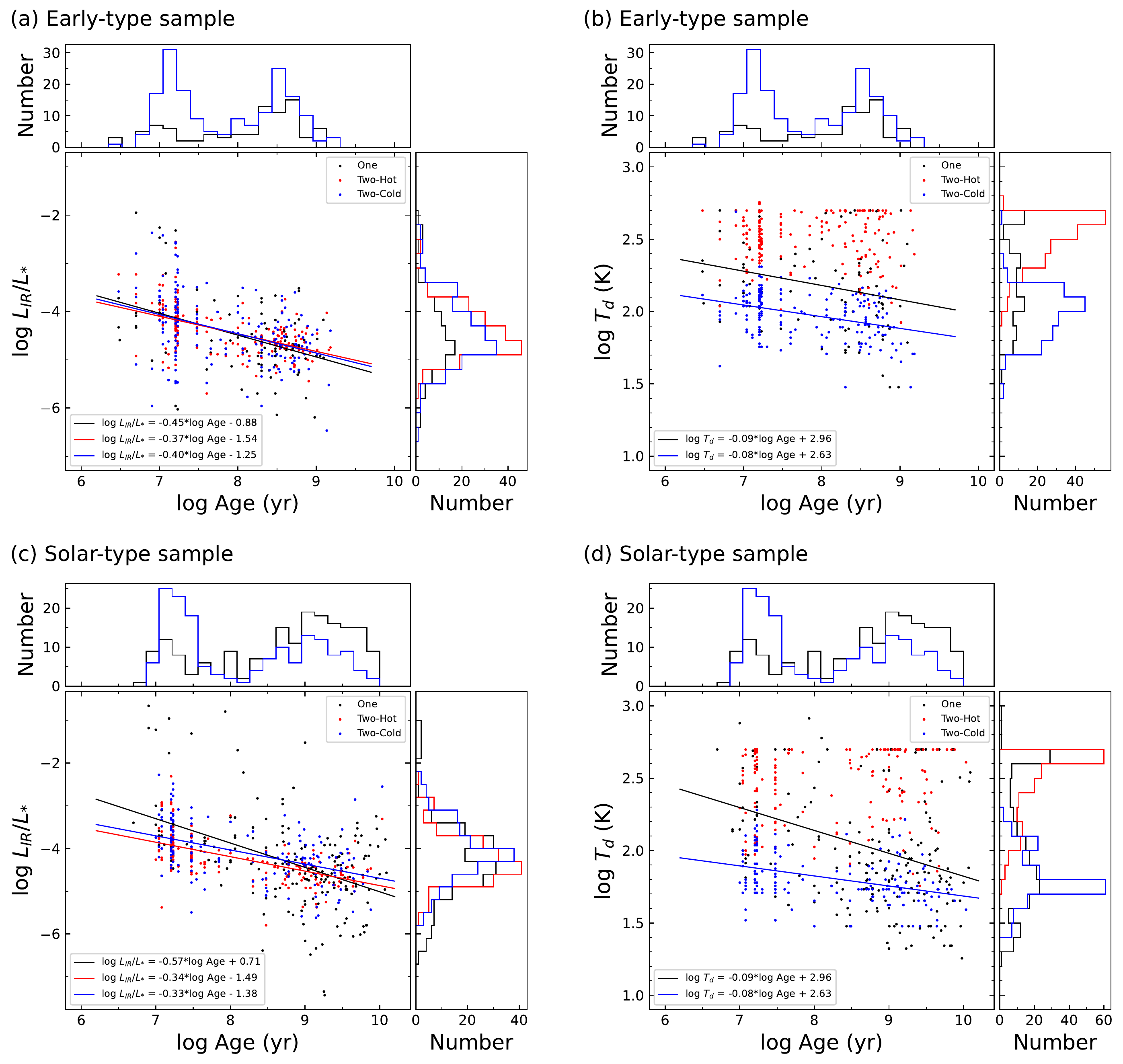}
	\caption{(a) Fractional luminosity and (b) dust temperature plotted as a function of stellar age for early-type sample. (c) Fractional luminosity and (d) dust temperature plotted as a function of stellar age for solar-type sample. Objects that are better fit using a one-temperature model are plotted as black circles. Objects that are better fit using a two-temperature model are plotted as red and blue circles, representing the warm and cold components, respectively.  The different colour lines are the best-fit trends to the data of corresponding colour sources.}
    	\label{figure3}
\end{figure}

On one hand, we study the fractional luminosity/dust temperature distribution with stellar age, as shown in Figure \ref{figure3}, where panels (a) and (b) represent the early-type sample and panels (c) and (d) represent the solar-type sample.
  From panels (a) and (c), we can find that, either in the early-type or solar-type sample, $L_{IR}/L_{*}$ decreases as a function of stellar age for dust in either one-belt systems or two-belt systems with one-belt systems decreasing faster as the black line shows. From panels (b) and (d), we also find that, either in the early-type or solar-type sample, $T_{d}$ for both one-belt systems and cold dust in two-belt systems decrease as a function of stellar age, consistent with dynamical evolution. While $T_{d}$ for the warm dust component in two-belt systems seems independent of stellar age. The fitting relationship can be seen from the figure.

\begin{figure}
	\centering
	\includegraphics[width=16cm,angle=0]{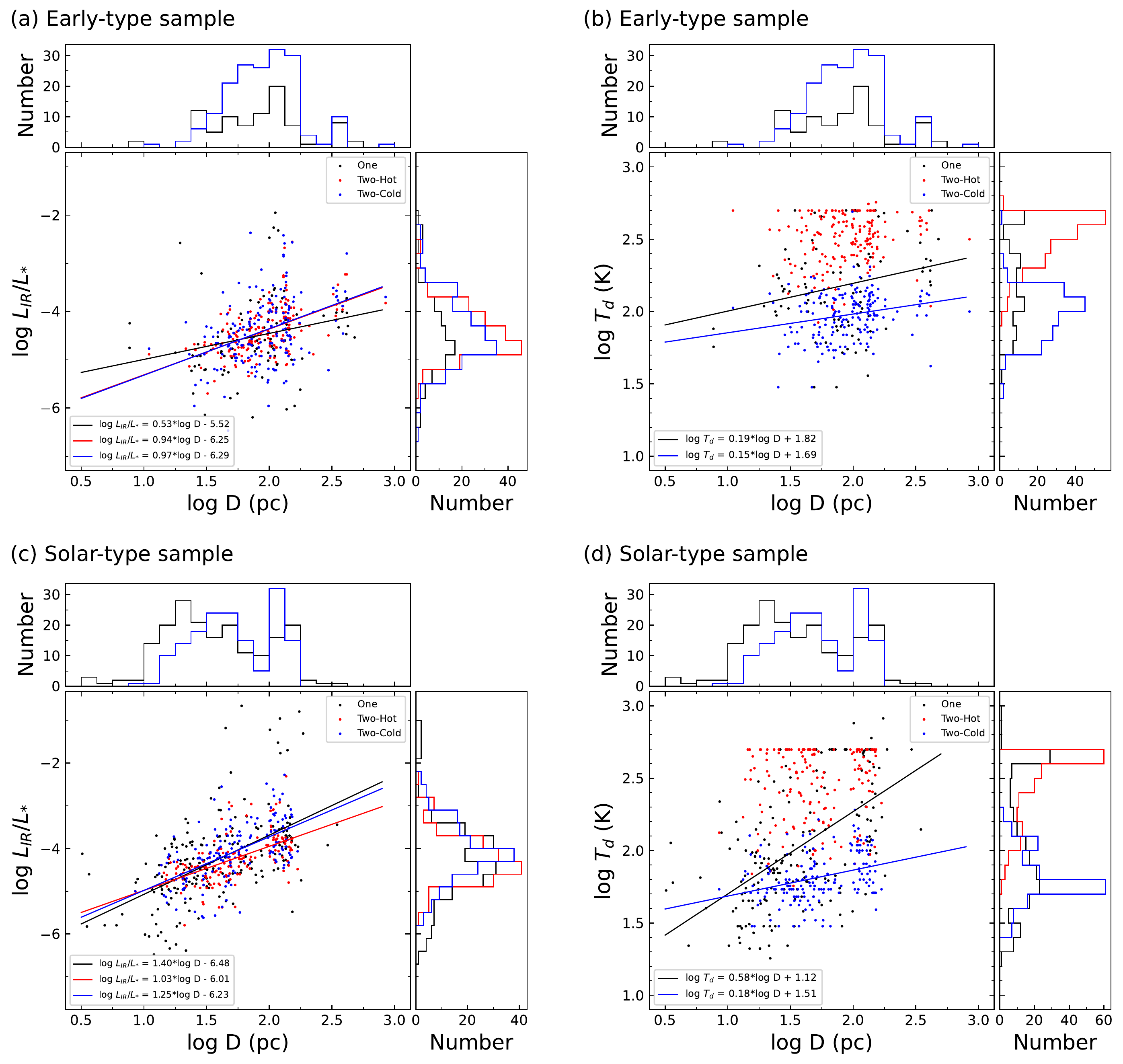}
	\caption{(a) Fractional luminosity and (b) dust temperature plotted as a function of distance for early-type sample. (c) Fractional luminosity and (d) dust temperature plotted as a function of distance for solar-type sample. Objects that are better fit using a one-temperature model are plotted as black circles. Objects that are better fit using a two-temperature model are plotted as red and blue circles, representing the warm and cold components, respectively. The different colour lines are the best-fit trends to the data of corresponding colour sources. }
	\label{figure4}
\end{figure}

On the other hand, we study the fractional luminosity/dust temperature distribution with distance, as shown in Figure \ref{figure4}, where panels (a) and (b) represent the early-type sample and panels (c) and (d) represent the solar-type sample.
From panels (a) and (c), we can find that $L_{IR}/L_{*}$ increase as a function of distance in both one-belt and two-belt systems in either early-type or solar-type sample. From panels (b) and (d), we can also find that, either in the early-type or solar-type sample, the dust temperature in one-belt systems and the temperature of cold dust in two-belt systems are positively correlated with distance as the black and blue line shows. While the temperature of warm dust in two-belt systems has no obvious correlation to the distance. The fitting relationship can be seen from the figure.

\section{Discussion}
\label{sect:discussion}

In the previous section, we study the properties and correlations between debris disks and their host stars. 
These dust properties we collected are mostly estimated from SED fitting in their original literature. While the interpretation of SEDs is ambiguous. So more detailed study of disk properties should need further observation such as resolved disks in scattered light or in the thermal emission from optical to millimeter wavelengths. 
And from the image, gas is also detected in many disks as well as planets. 
Therefore, we also collected the resolved information of our catalog disks as well as the planet and gas information. 
In this section, we intend to discuss the distribution difference among these disks and search for the correlations between disks and planets. 

\subsection{Observational classification: gas-detected, planet-detected and resolved disks}

According to the observation properties, we classified the total sample into 3 groups: gas-detected, planet-detected and resolved disks.
The resolved debris disks were collected from the Catalog of Circumstellar Disk\footnote{\it http://www.circumstellardisks.org} and there were 103 disks in total. The gas-detected debris disks were collected from the sample of our previous work \citep{2022AcASn..63...69C}
and there were 37 disks. 
We cross-matched our total sample of 1095 stars to 
The Extrasolar Planets Encyclopaedia \footnote{http://exoplanet.eu/}
and got 73 stars with planets.
We collected these three groups of disks and marked them in the first column of Table \ref{table:1}: g represents the gas, r represents resolved and p represents the planet. 
While we only study the 612 sources in the prime debris disks sample in the previous section, in this section we also focus on this sample. Among these 612 sources, there are 96 resolved disks, 47 with planets and 31 with gas detected.

\begin{figure}
	\centering
     \includegraphics[width=16cm,angle=0]{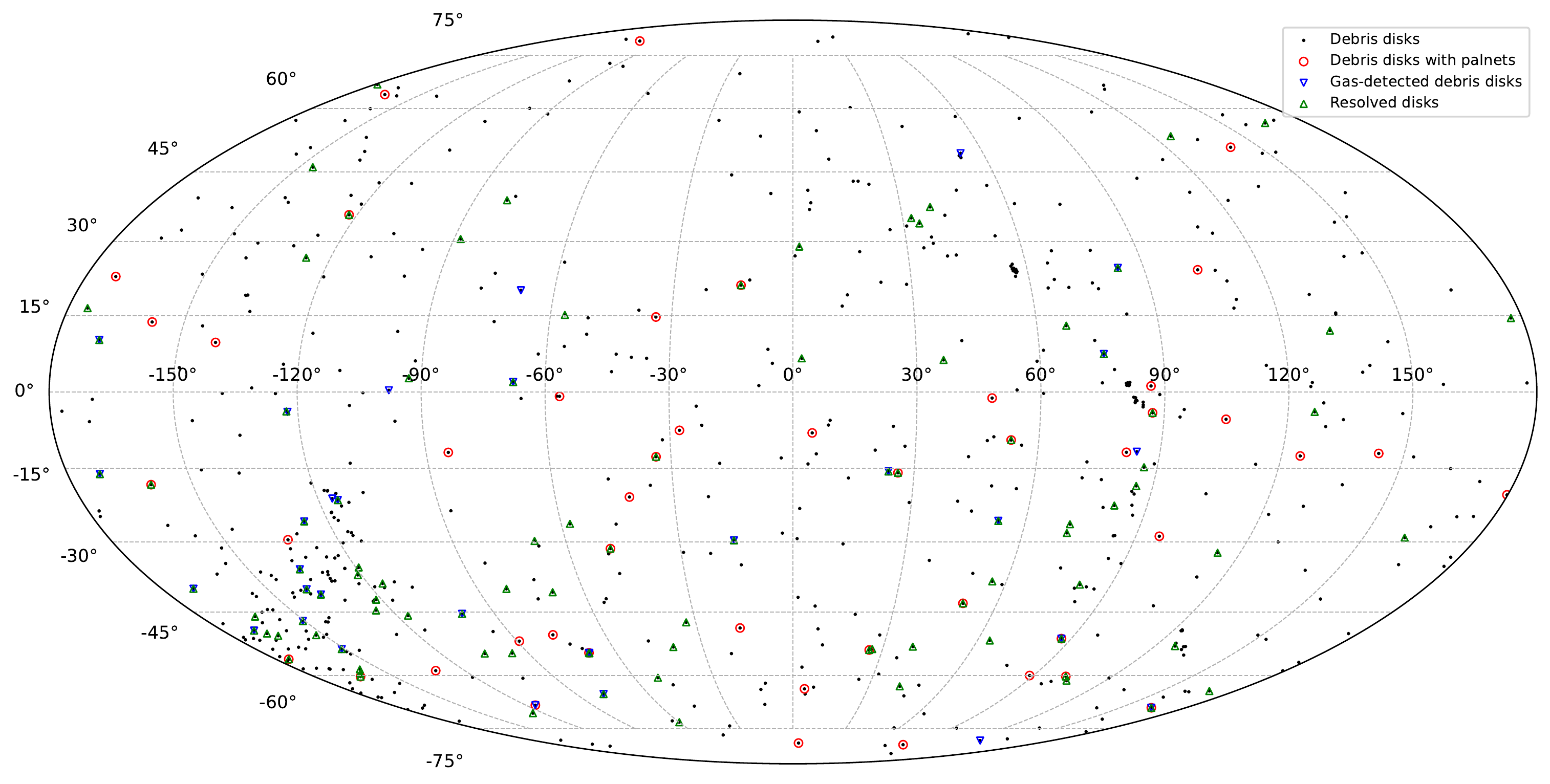}
	\caption{The sky distribution of debris disks, shown in Equatorial coordinates. Objects are over-plotted as red circles, blue inverted triangles, and green triangles, representing the disks with planets, gas-detected disks, and resolved disks, respectively.}
	\label{figure5}
\end{figure}

Next, We search for the distribution difference between different groups.
Firstly, we can see their distribution on the celestial sphere as Figure \ref{figure5} shows. 
All groups of debris disks are more distributed in the southern sky: 387/612 debris disks, 36/47 disks with planets, 24/31 gas-detected disks, and 71/96 resolved disks. 

\begin{figure}[h]
	\centering
     \includegraphics[width=10cm,angle=0]{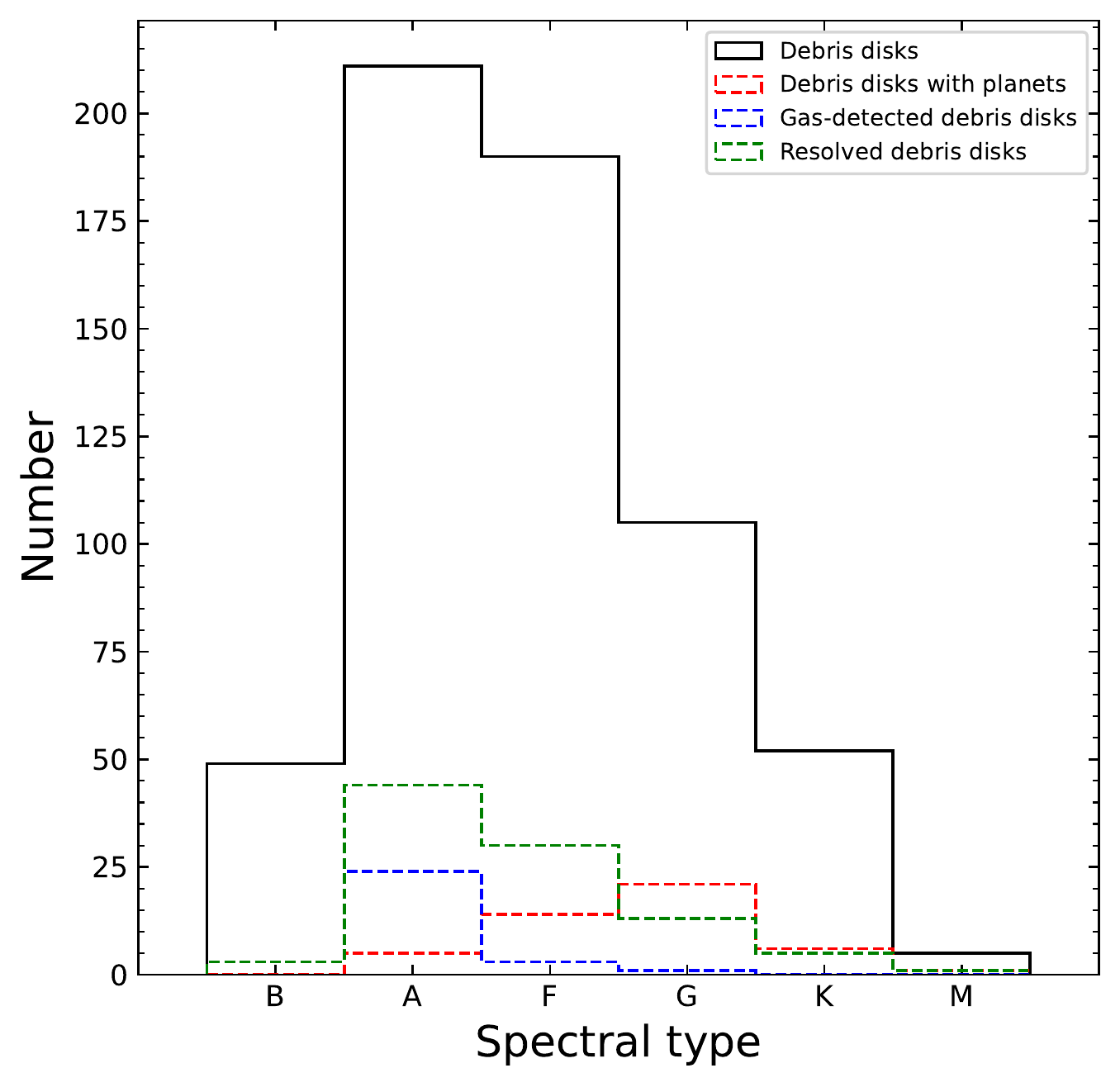}
	\caption{Spectral type histogram showing the number of targets observed as a function of spectral type. The 612 sources in debris disk sample are plotted as a black line. Objects are plotted as red dash line,  blue dash line, and green dash line, representing the 47 disks with planets, 31 gas-detected disks, and 96 resolved disks, respectively. }
	\label{figure6}
\end{figure}

Secondly, we can see the spectral type distributions between different groups in Figure \ref{figure6}: the resolved disks group has a similar distribution to the debris disk sample with mostly A, F, and G-type stars;
debris disks detected with planets are mostly solar-type stars; gases are easier to detect in the disks of early-type stars.

\begin{figure}[h]
	\centering
	\includegraphics[width=10cm,angle=0]{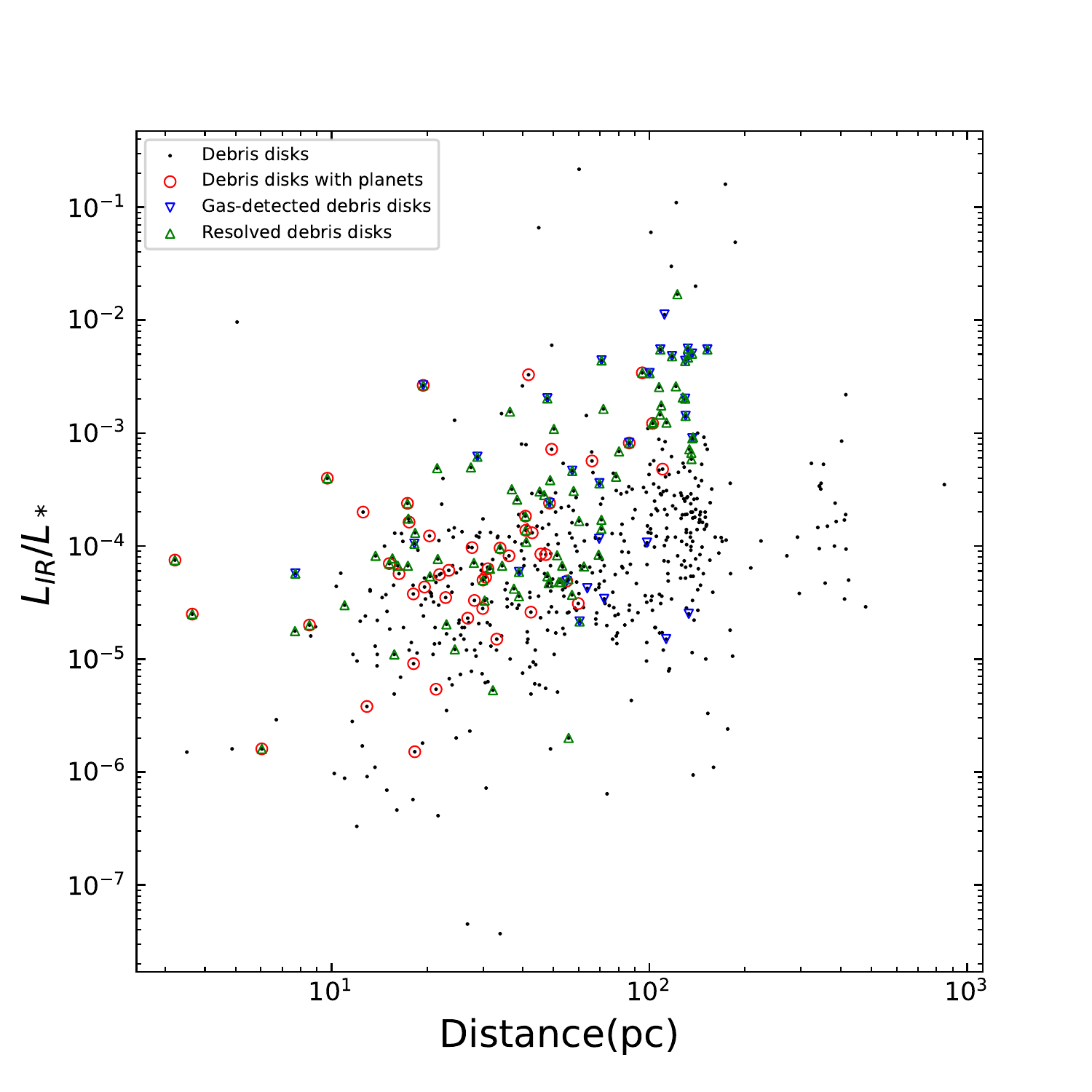}
	\caption{Fractional luminosity, $L_{IR}/L_{*}$, versus distance from Earth. Objects are over-plotted as red circles, blue inverted triangles, and green triangles, representing the disks with planets, gas-detected disks, and resolved disks, respectively.}
	\label{figure7}
\end{figure}

Thirdly, we plot fractional luminosity $L_{IR}/L_{*}$ as a function of distance in Figure \ref{figure7}. 
We find that these three groups have no obvious different distribution in distance, mostly located within 150 pc. While the fractional luminosity has different distribution with gas-detected debris disks tends to be higher than disks with planets.

\begin{figure}[h]
	\centering
	\includegraphics[width=10cm,angle=0]{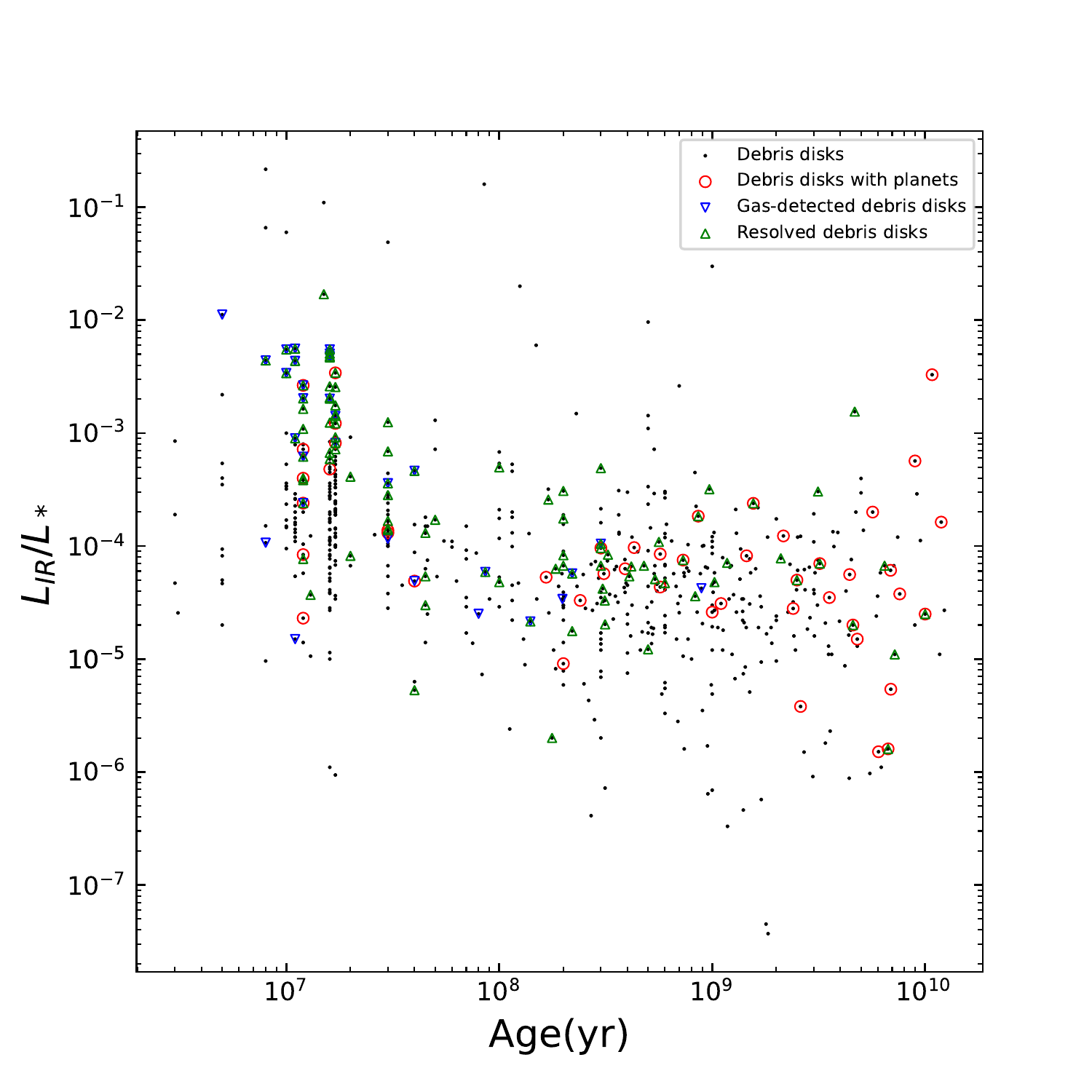}
	\caption{Fractional luminosity,${L_{IR}/L_{*}}$, versus stellar age. Objects are over-plotted as red circles, blue inverted triangles, and green triangles, representing the disks with planets, gas-detected disks, and resolved disks, respectively. }
	\label{figure8}
\end{figure}

Lastly, we plot fractional luminosity $L_{IR}/L_{*}$ as a function of stellar age in Figure \ref{figure8}. 
We find that gases are mostly detected around young disks while planets are mostly found around older stars which represents the evolution of debris disks and planetary systems. 

\subsection{Correlations between debris disks and planets}
From the above distributions, we can see the evolution trend of debris disks and planets.  
Debris disks and planets are believed to form in protoplanetary disks and debris disks can be used as a tool to search for exoplanet systems. But until now, the correlation between exoplanets and debris disks has remained inconclusive.

With a larger sample, we intend to investigate more correlations between planets and debris disks. An effective method is to follow and compare with previous works. One typical sample is Maldonado’s sample which has 29 debris disks with planets \citep{2012A&A...541A..40M}. We will refer to this sample as Maldonado12 sample, hereafter. Among the 29 stars, 11 stars host known multiplanet systems, which represents an incidence rate of 38\%. In contrast, our debris disks sample have 73 disks with planets of which 29 host multiplanet systems, which represents an slightly higher incidence rate of 40\%.
While the properties collected from The Extrasolar Planets Encyclopaedia are incomplete (there are many planets without eccentricity or orbit semi-major axis), we also searched for other exoplanet websites \footnote{\it https://exoplanets.nasa.gov/} 
\footnote{\it http://www.exoplanetkyoto.org/} and collected the planet properties into Table \ref{table:3}, including the name of host stars, the name, mass, orbit semi-major axis and eccentricity of planets. Next, we will discuss the distribution differences and correlations between disks and planets.

\begin{figure}
	\centering
	\includegraphics[width=10cm,angle=0]{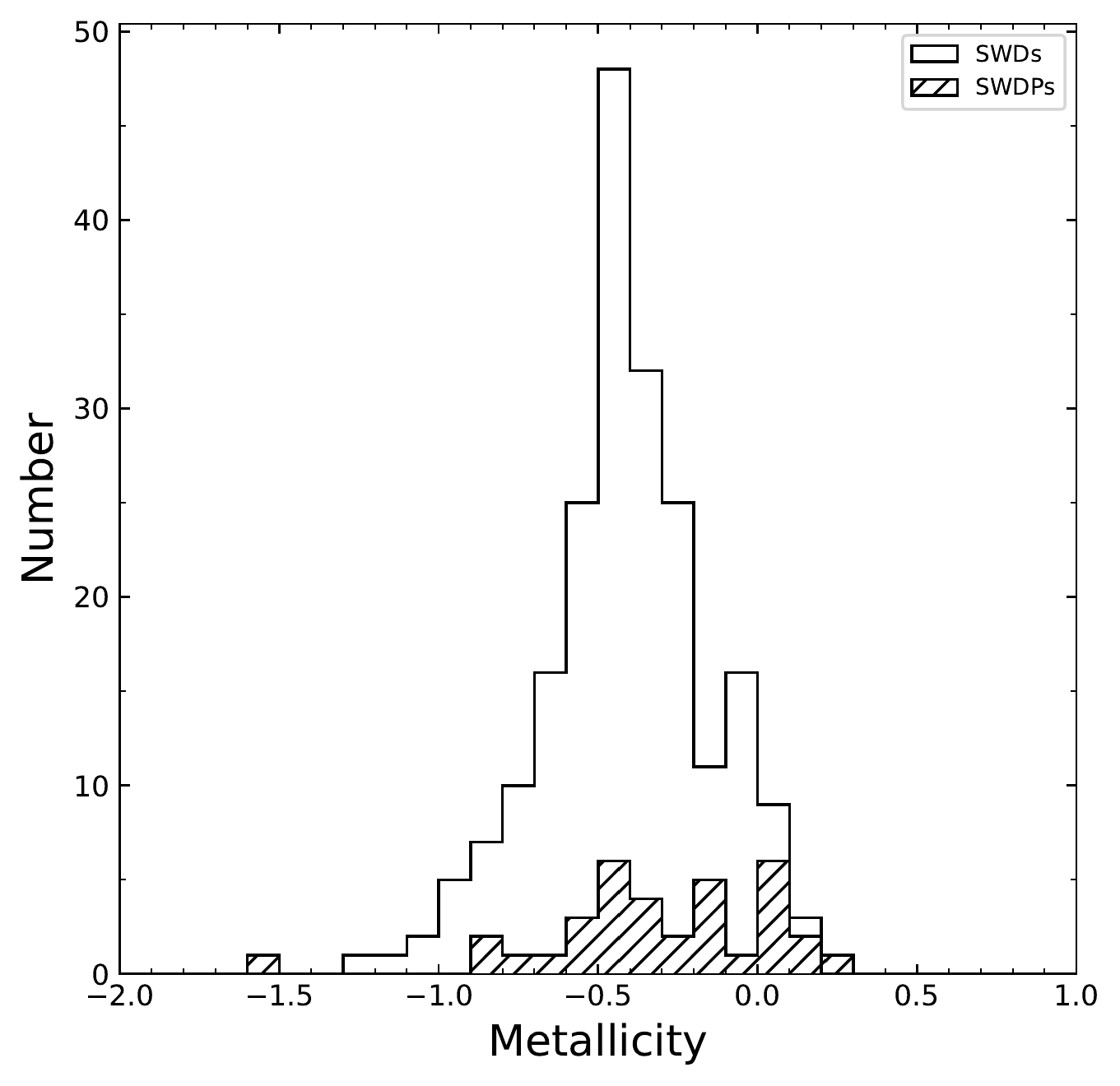}
	\caption{Metallicity histogram showing the number of targets observed as a function of metallicity.}
	\label{figure9}
\end{figure}

On one hand, we compare the distribution of stars with debris disks but no planets (SWDs) and stars with debris disks and planets (SWDPs). From Figures \ref{figure7} and \ref{figure8}, we can see there are different distributions between SWDs and SWDPs in terms of age and distance, with the SWDs containing more distant stars and younger stars. In order to avoid the possible bias from more distant and younger stars, we cut the SWDs to the same distance and age range of SWDPs with ages from 10Myr to 10Gyr and distances within 120 pc. At the same time, in order to avoid bias from wide spectral types, we only study solar-type stars. After that, we define a sample (hereafter planet sample) with 211 SWDs and 35 SWDPs with metallicity information. From Figure \ref{figure9}, we can see that it is a different distribution of metallicity between SWDs and SWDPs with $P<0.05$ of the Kolmogorov-Smirnov (K-S) test. Our results suggest that SWDPs (average = -0.31) have higher metallicities than SWDs (average = -0.41). 

On the other hand, we search for the correlations between disks and different types of planets. We use the same criterion as previous papers \citep{2009ApJ...693.1084W, 2012A&A...541A..40M, 2015ApJ...801..143M}
 to classify planets: the mass limit is 30${M_ \oplus}$ for the low-mass planet and Jupiter; the hot and cold planets are divided by a planet semi-major axis of 0.1AU.
In our planet sample, 35 SWDPs are divided into four groups: cool low-mass planets, hot low-mass planets, cool Jupiters, and hot Jupiters, as shown in the last column of Table \ref{table:3}. 
There are 30 SWDPs host cool Jupiters and 6 stars host the other 3 types of planets. Note there is a star host both cool Jupiter and low-mass planets.
Next, we will discuss distributions according to planetary classification.

\begin{figure}
	\centering
	\includegraphics[width=10cm,angle=0]{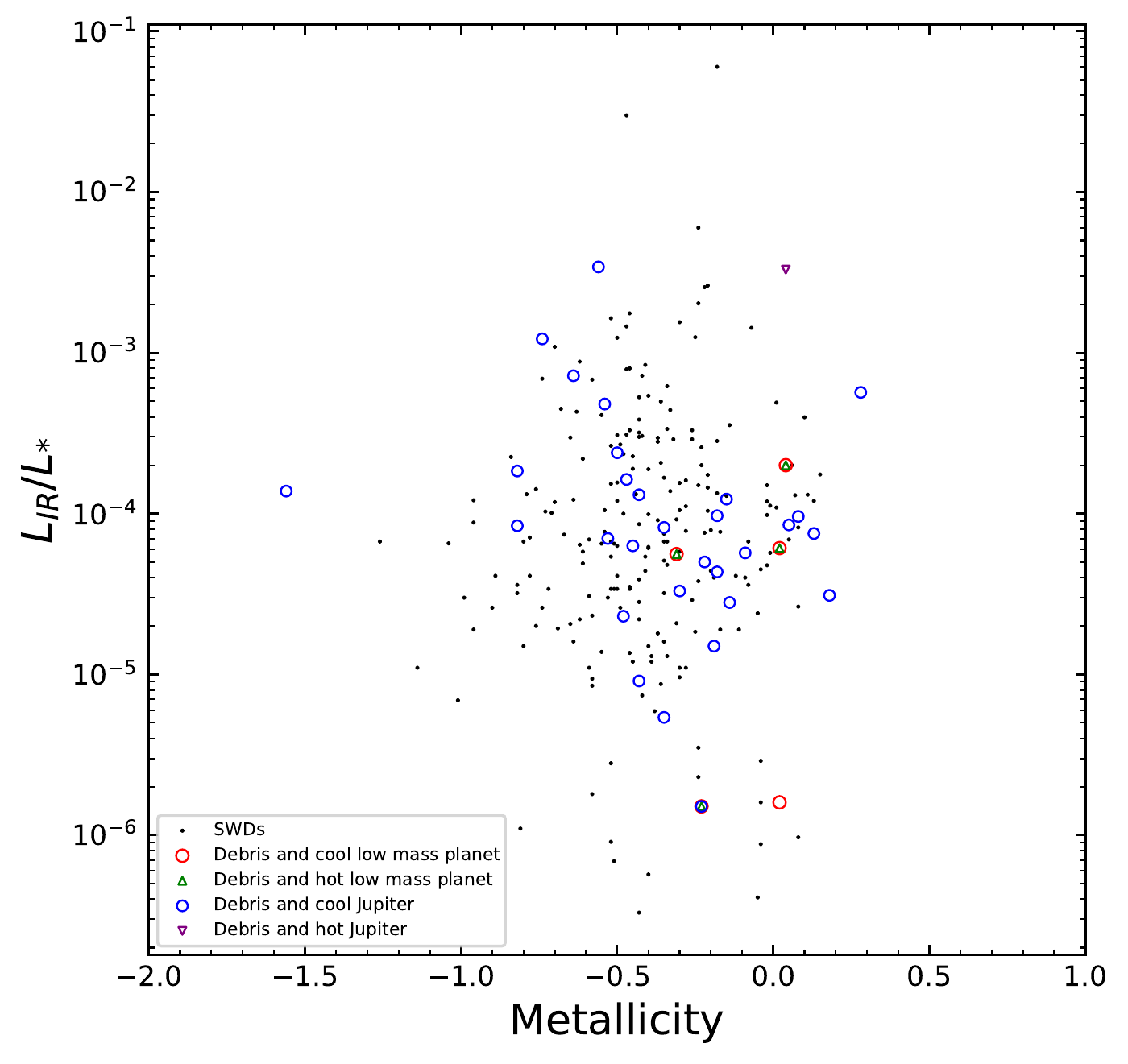}
	\caption{Fractional luminosity, $L_{IR}/L_{*}$, versus metallicity for those stars hosting debris disks. Objects are over-plotted as red circles, green triangles, blue circles, and purple inverted triangles, representing cool low-mass planets, hot low-mass planets, cool Jupiters, and hot Jupiters, respectively. }
	\label{figure10}
\end{figure}

Firstly, we search for the correlation between fractional luminosity and metallicity as shown in Figure \ref{figure10}. We find the fractional luminosity of SWDPs approximately distribute in two orders of magnitude from ${10^{-3}}$ to  ${10^{-5}}$ and are well-mixed with SWDs. The metallicities of stars with disks and cool Jupiter are widely distributed from -1.56 to 0.28, while the other 3 groups fall on the higher metallicity zone ($\textgreater$ -0.3).
Furthermore, we find more than half of SWDPs are concentrated in the low-fractional luminosity/high-metallicity corner of Figure \ref{figure10}, which is consistent with the metallicity trend in Maldonado12 sample. 

\begin{figure}
	\centering
	\includegraphics[width=10cm,angle=0]{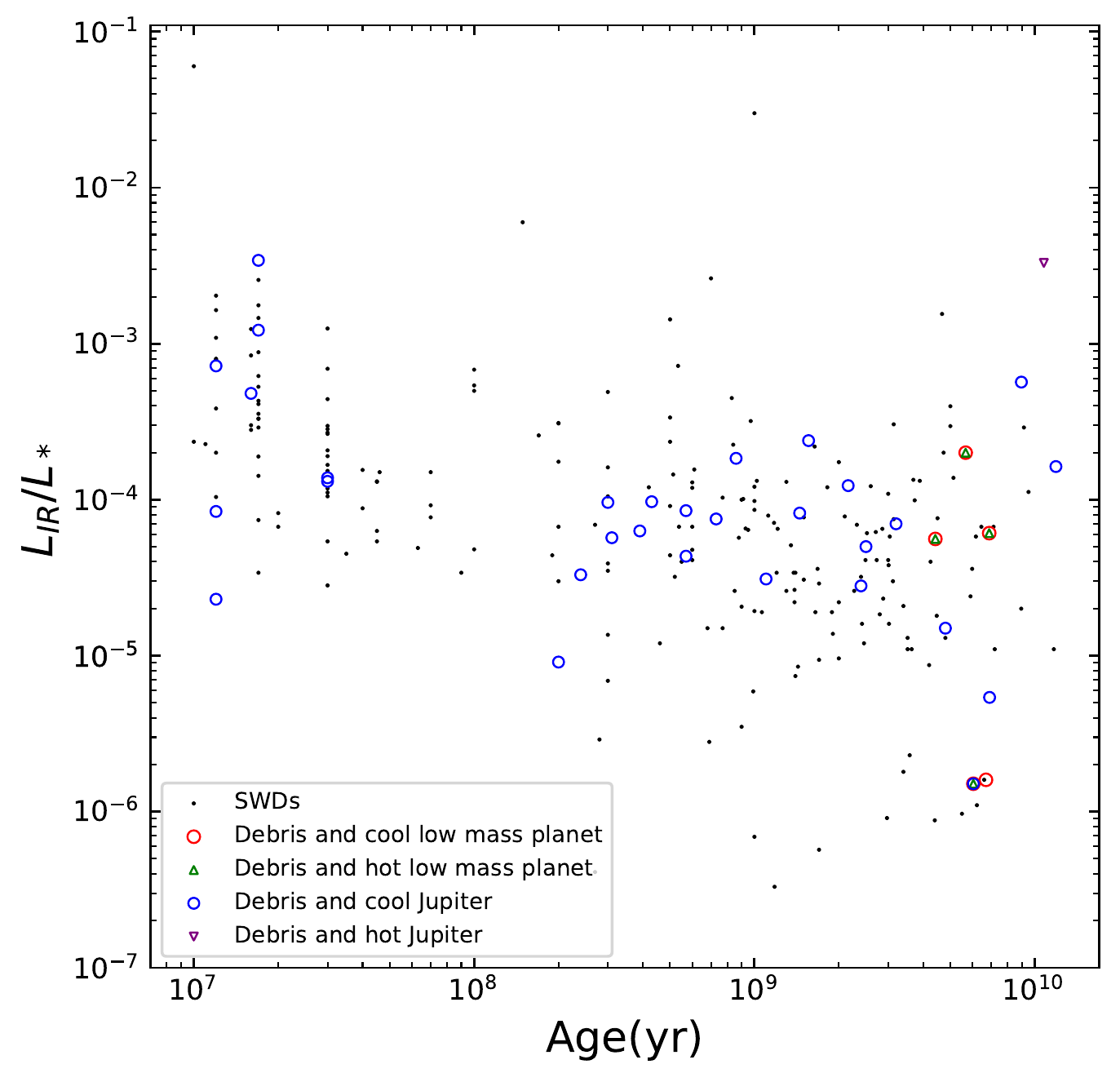}
	\caption{Fractional luminosity, ${L_{IR}/L_{*}}$, versus stellar age. Objects are over-plotted as red circles, green triangles, blue circles, and purple inverted triangles, representing cool low-mass planets, hot low-mass planets, cool Jupiters, and hot Jupiters, respectively.}
	\label{figure11}
\end{figure}

Secondly, we search for the correlation between fractional luminosity and stellar age as Figure \ref{figure11} shows. The SWDPs and SWDs in our planet sample have the same range of age, distance, and spectral type, however, unlike SWDs widely distributed from 10 Myr to 10 Gyr, most SWDPs (27 in 35) are older than 100 Myr. The ages of stars with disks and cool Jupiter are widely distributed from 10 Myr to 10 Gyr, while the other 3 groups fall in the larger age zone ($\textgreater$ 4 Gyr). 

\begin{figure}
	\centering
	\includegraphics[width=16cm,angle=0]{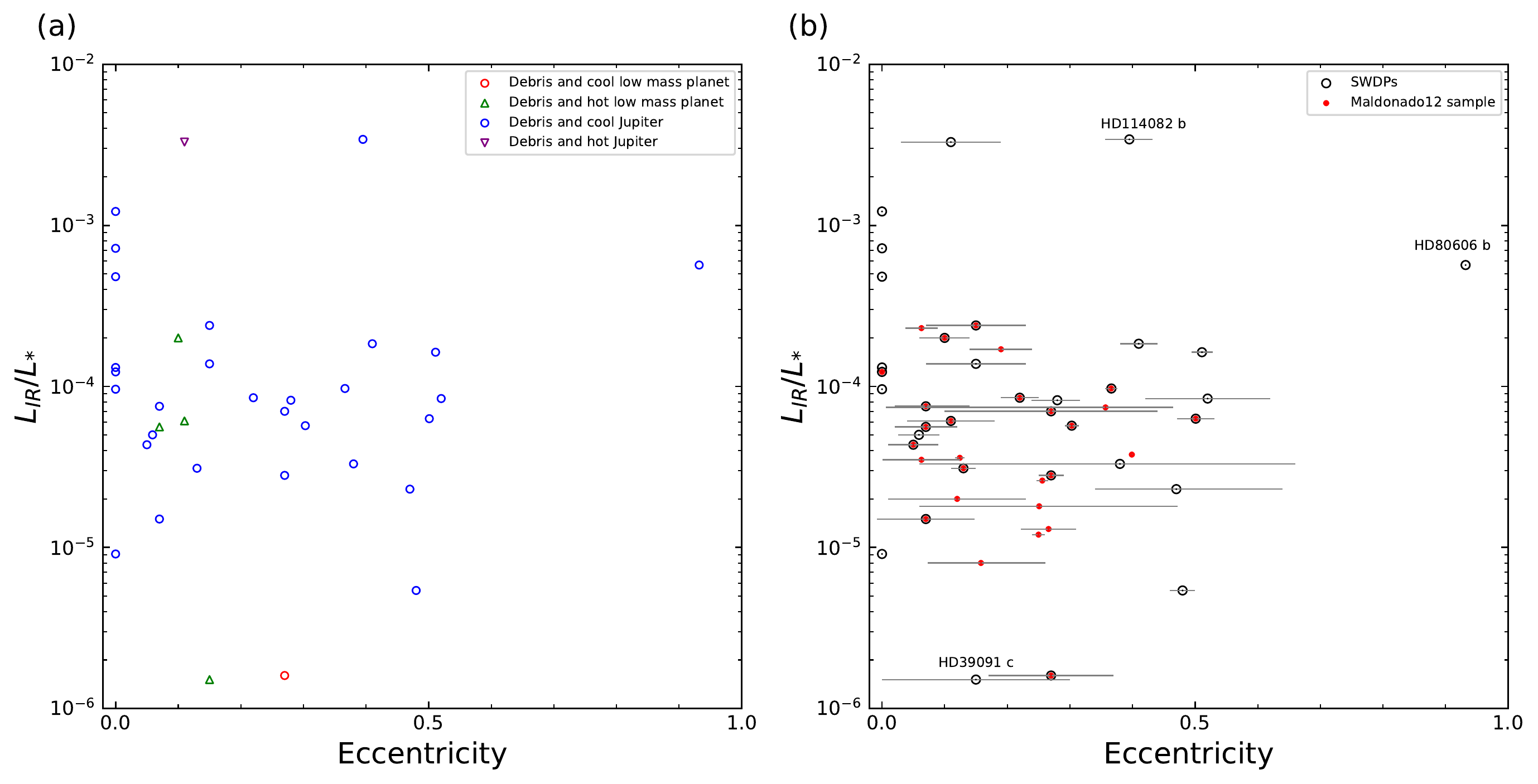}
	\caption{Fractional luminosity, ${L_{IR}/L_{*}}$, versus eccentricity. (a) Comparison of different types of planets in our planet sample. Objects are plotted as red circles, green triangles, blue circles, and purple inverted triangles, representing cool low-mass planets, hot low-mass planets, cool Jupiters, and hot Jupiters, respectively. (b) Comparison of Maldonado12 sample with our SWDPs in planet sample. Objects are plotted as black circles and red dots, representing SWDPs and Maldonado12 sample, respectively.}
	\label{figure12}
\end{figure}

Finally, we search for the correlation between the fractional luminosity and eccentricity (we take as reference the innermost planet), and compare it with Maldonado12 sample, as Figure \ref{figure12} shows. Our 35 SWDPs are shown in panel (a) with 4 groups of planets. The eccentricities of cool Jupiters are widely distributed from 0 to 0.932, while the other 3 groups fall on the smaller eccentricity zone ($\textless$ 0.3).
Then we compare our sample to Maldonado12 sample as panel (b) shows. The uncertainty of eccentricity is over plot to each planet but not the uncertainty of fractional luminosity which was not given in the original papers. There are 16 common sources in Maldonado12 sample and our SWDPs. The remaining stars in Maldonado12 sample are in our debris disk total sample but have been removed from the planet sample due to the sample selection. We find most SWDPs in our planet sample meet the trend in Maldonado12 sample that the luminosity of the dust decreases with larger eccentricity. Neverthless, there are three exceptions which are HD39091 c (lower left), HD114082 b (top middle) and HD80606 b (top right), it can be alternatively explained with large eccentricity uncertainties for HD39031 c with 0.15 $\pm$ 0.15 while it does not work for the other two.
Note for HD80606 b, the very eccentric orbit (0.93183 $\pm$ 0.00014) of this planet is probably due to the influence of its binary star HD80607 \citep{2003ApJ...589..605W} which locates 20.6$\arcsec$ away.

\section{Summary}

This paper aims to supply a catalog of 1095 debris disks collected from published literature and define a less biased debris disk sample with 612 sources from the catalog.  

We collected the stellar and disk properties from SIMBAD, Gaia DR3 and referenced literature. Then we studied the distributions of properties and correlations between debris disks and their host stars. Firstly, from the distributions of stellar properties, we found that debris disks were widely distributed from B to M-type stars with most located within 200 pc.
Secondly, we showed the distributions of disk properties in two groups with 281 one-belt systems and 331 two-belt systems. We found that the fractional luminosities of one-belt systems were widely distributed from $10^{-7}-10^{-1}$ and two-belt systems were narrowly distributed from $10^{-6}-10^{-3}$, while there was no obvious difference between cold and warm components in two-belt systems. While there were no obvious different distributions of dust temperatures between one-belt and two-belt systems. 
Thirdly, we searched for the correlations between disks and stars in two subsamples: solar-type sample and early-type sample. 
Either in early-type or solar-type sample, the fractional luminosity dropped off with age and the dust temperature of both one-belt systems and cold components in two-belt systems decreased with stellar age too. 
And either in early-type or solar-type sample, the fractional luminosity increased with stellar distance in both one-belt and two-belt systems, and the dust temperature in one-belt systems and the cold dust temperature in two-belt systems was positively correlated with distance too.

Furthermore, in order to know more about the disk component and structure, it is essential to further observe them through direct imaging. 
And from the image, gas and/or planet were also detected in some of the disks. As a result, we also collected the resolved information in our catalog as well as the planet and gas information. There are 96 resolved disks, 47 disks with planets and 31 disks with gas detected in our debris disk sample.
After that, we discussed the distribution difference among these three groups.
All debris disks in either group were more distributed in the southern sky and almost all disks in these three groups were located within 150 pc. Planets were mostly found around solar-type stars, gases were easier to detect around early-type stars and resolved disks were mostly distributed from A to G-type stars. 
The fractional luminosity had different distributions with gas-detected debris disks tended to be higher than disks with planets.
Planets were mostly found around old stars and gas-detected disks were much younger which represented the evolution of debris disks and planetary systems. 

Last but not least, we searched for the correlations between disks and planets. In order to avoid the possible bias from the stellar distance, age and spectral type, we defined a planet sample with 211 SWDs and 35 SWDPs. We found SWDPs had higher metallicities than SWDs with more than half of the SWDPs concentrated in the low-fractional luminosity/high-metallicity corner. Moreover, we searched for the correlations between disks and different types of planets. There were 30 SWDPs hosted cool Jupiters and 6 stars hosted the other three types of planets. 
We found the stars with disks and cold Jupiters were distributed over a wide age (10 Myr-10 Gyr) and metallicity (-1.56-0.28) range while the other three groups were located on the old ($\textgreater$ 4Gyr) and metal-rich ($\textgreater$ -0.3) regions. The fractional luminosities of the majority sources (33 in 35) tended to decrease with the increase of eccentricities of planets. And the eccentricities of cool Jupiters (0 to 0.932) were distributed wider than the other three types of planets ($\textless$ 0.3). 

\begin{acknowledgements}

This research has made use of the SIMBAD database, operated at CDS, Strasbourg, France. 
This work presents results from the European Space Agency (ESA) space mission Gaia. Gaia data are being processed by the Gaia Data Processing and Analysis Consortium (DPAC). Funding for the DPAC is provided by national institutions, in particular the institutions participating in the Gaia MultiLateral Agreement (MLA).
This research has made use of the NASA Exoplanet Archive, which is operated by the California Institute of Technology, under contract with the National Aeronautics and Space Administration under the Exoplanet Exploration Program.

\end{acknowledgements}

%\begin{thebibliography}{99}
%\bibitem[Abraham et al.(1999)]{abr99} Abraham, P., Leinert, C., Burkert, A., Lemke, D., Henning, T. 1999, \aap, 338, 91
  
%\end{thebibliography}
\bibliographystyle{raa}
\bibliography{bibref}

\clearpage
	\begin{longtable}{lllllc}
		\caption{\label{table:3}Stars with known debris disks and planets. }\\
		\hline
	Host Name   & Planet Name  & M$_{p}$  &  a  &   e  & Classification  \\
	&                & (${M_ \oplus}$)     & (AU)  &    &   \\
		\hline		
		HD 142     & HD 142 A b       & 395.5   & 1.038    & 0.158 & cj \\
           & HD 142 A c       & 3521.3  & 9.815    & 0.277 & cj \\
HD 1237A   & GJ 3021 b        & 1071.0  & 0.490    & 0.511 & cj \\
HD 1461    & HD 1461 b        & 6.4     & 0.063    & 0.110 & hl \\
           & HD 1461 c        & 5.6     & 0.112    & 0.305 & cl \\
HD 1466    & HIP 1481 b       & 953.0   & 2.000    & 0 & cj \\
HD 10647   & HD 10647 b       & 298.8   & 2.015    & 0.150 & cj \\
HD 10700   & tau Cet g        & 1.8     & 0.133    & 0.060 & cl \\
           & tau Cet h        & 1.8     & 0.243    & 0.230 & cl \\
           & tau Cet e        & 3.9     & 0.538    & 0.180 & cl \\
           & tau Cet f        & 3.9     & 1.334    & 0.160 & cl \\
HD 19994   & HD 19994 A b     & 435.4   & 1.305    & 0.063 & cj \\
HD 20794   & HD 20794 b       & 2.7     & 0.121    & 0.270 & cl \\
           & HD 20794 d       & 4.8     & 0.350    & 0 & cl \\
           & HD 20794 e       & 4.8     & 0.509    & 0.290 & cl \\
HD 22049   & eps Eridani b    & 209.8   & 3.530    & 0.070 & cj \\
HD 29391   & 51 Eri b         & 635.7   & 13.200   & 0.450 & cj \\
HD 33564   & HD 33564 b       & 2892.1  & 1.100    & 0.340 & cj \\
HD 35850   & AF Lep b         & 1621.0  & 8.400    & 0.470 & cj \\
HD 38529   & HD 38529 b       & 206.3   & 0.115    & 0.256 & cj \\
           & HD 38529 c       & 3320.5  & 3.225    & 0.357 & cj \\
HD 38858   & HD 38858 b       & 30.6    & 1.038    & 0.270 & cj \\
HD 39060   & beta Pic c       & 2860.5  & 2.700    & 0.240 & cj \\
           & beta Pic b       & 3496.1  & 9.100    & 0.080 & cj \\
HD 39091   & pi Men c         & 3.6     & 0.069    & 0.150 & hl \\
           & pi Men b         & 3890.0  & 3.309    & 0.642 & cj \\
           & pi Men d         & 13.4    & 0.503    & 0.220 & cl \\
HD 40307   & HD 40307 b       & 4.0     & 0.047    & 0.200 & hl \\
           & HD 40307 c       & 6.6     & 0.080    & 0.060 & hl \\
           & HD 40307 d       & 9.5     & 0.132    & 0.070 & cl \\
           & HD 40307 e       & 3.5     & 0.189    & 0 & cl \\
           & HD 40307 f       & 5.2     & 0.247    & 0.020 & cl \\
           & HD 40307 g       & 7.1     & 0.600    & 0.290 & cl \\
HD 40979   & HD 40979 b       & 1484.3  & 0.850    & 0.250 & cj \\
HD 44627   & AB Pic b         & 4290.5  & 260.000  & 0 & cj \\
HD 45184   & HD 45184 b       & 12.2    & 0.064    & 0.070 & hl \\
           & HD 45184 c       & 8.8     & 0.110    & 0.070 & cl \\
HD 46375A  & HD 46375 A b     & 71.8    & 0.040    & 0.063 & hj \\
HD 50499   & HD 50499 b       & 520.0   & 3.833    & 0.266 & cj \\
           & HD 50499 c       & 931.2   & 9.020    & 0 & cj \\
HD 50554   & HD 50554 b       & 1574.5  & 2.353    & 0.501 & cj \\
HD 50571   & HR 2562 b        & 9534.9  & 20.300   & 0 & cj \\
HD 52265   & HD 52265 b       & 384.6   & 0.520    & 0.270 & cj \\
HD 69830   & HD 69830 b       & 10.2    & 0.079    & 0.100 & hl \\
           & HD 69830 c       & 11.8    & 0.186    & 0.130 & cl \\
           & HD 69830 d       & 18.1    & 0.630    & 0.070 & cl \\
HD 73526   & HD 73526 b       & 978.9   & 0.650    & 0.190 & cj \\
           & HD 73526 c       & 715.1   & 1.030    & 0.280 & cj \\
HD 75732   & 55 Cnc e         & 8.0     & 0.015    & 0.050 & hl \\
           & 55 Cnc b         & 264.0   & 0.113    & 0 & cj \\
           & 55 Cnc c         & 54.5    & 0.237    & 0.030 & cj \\
           & 55 Cnc d         & 1232.5  & 5.957    & 0.130 & cj \\
           & 55 Cnc f         & 44.8    & 0.771    & 0.080 & cj \\
HD 80606   & HD 80606 b       & 1392.1  & 0.457    & 0.932 & cj \\
HD 82943   & HD 82943 c       & 622.6   & 0.743    & 0.366 & cj \\
           & HD 82943 b       & 534.3   & 1.183    & 0.162 & cj \\
           & HD 82943 d       & 92.2    & 2.145    & 0 & cj \\
HD 95086   & HD 95086 b       & 1589.0  & 55.700   & 0.200 & cj \\
HD 104067  & HD 104067 b      & 50.9    & 0.260    & 0 & cj \\
HD 106252  & HD 106252 b      & 3178.0  & 2.610    & 0.480 & cj \\
HD 106906  & HD 106906 (AB) b & 3496.0  & 650.000  & 0 & cj \\
HD 108874  & HD 108874 b      & 451.3   & 1.040    & 0.130 & cj \\
           & HD 108874 c      & 314.6   & 2.659    & 0.239 & cj \\
HD 113337  & HD 113337 b      & 1032.0  & 1.021    & 0.280 & cj \\
           & HD 113337 c      & 6269.0  & 6.828    & 0.164 & cj \\
HD 114082  & HD 114082 b      & 2543.0  & 0.511    & 0.395 & cj \\
HD 114613  & HD 114613 b      & 113.5   & 5.340    & 0.458 & cj \\
HD 115617  & 61 Vir b         & 5.1     & 0.050    & 0.120 & hl \\
           & 61 Vir c         & 18.2    & 0.218    & 0.140 & cl \\
           & 61 Vir d         & 22.9    & 0.476    & 0.350 & cl \\
HD 117176  & 70 Vir b         & 2380.5  & 0.481    & 0.399 & cj \\
HD 128311  & HD 128311 b      & 562.2   & 1.084    & 0.303 & cj \\
           & HD 128311 c      & 1204.3  & 1.740    & 0.159 & cj \\
HD 131496  & HD 131496 b      & 572.0   & 2.010    & 0.181 & cj \\
HD 133803  & HIP 73990 b      & 6674.0  & 20.000   & 0& cj \\
           & HIP 73990 c      & 6992.0  & 32.000   & 0 & cj \\
HD 134060  & HD 134060 b      & 10.1    & 0.044    & 0.450 & hl \\
           & HD 134060 c      & 29.3    & 2.393    & 0.110 & cl \\
HD 134319  & TOI-1860 b       & 2.3     & 0.020    & 0 & hl \\
HD 135379  & beta Cir b       & 17798.0 & 6656.000 & 0 & cj \\
HD 137759  & HIP 75458 b      & 3756.7  & 1.453    & 0.701 & cj \\
           & HIP 75458 c      & 4958.1  & 19.400   & 0.455 & cj \\
HD 141004  & HIP 77257 b      & 13.6    & 0.124    & 0.160 & cl \\
HD 142091  & kappa CrB b      & 635.7   & 2.760    & 0.059 & cj \\
HD 145689  & HIP 79797 Ba     & 18434.0 & 350.000  & 0 & cj \\
           & HIP 79797 Bb     & 17841.0 & 350.000  & 0 & cj \\
HD 150706  & HD 150706 b      & 861.3   & 6.700    & 0.380 & cj \\
HD 160691  & mu Ara d         & 10.2    & 0.093    & 0.067 & hl \\
           & mu Ara b         & 1366.7  & 1.500    & 0.128 & cj \\
           & mu Ara c         & 1398.4  & 4.170    & 0.099 & cj \\
           & mu Ara e         & 2224.8  & 0.934    & 0.067 & cj \\
HD 164249  & HIP 88399 b      & 1907.0  & 8.000    & 0 & cj \\
HD 168746  & HD 168746 b      & 85.8    & 0.070    & 0.110 & hj \\
HD 169142  & HD 169142 b      & 953.0   & 37.200   & 0 & cj \\
HD 174429  & PZ Tel b         & 20341.0 & 20.000   & 0.520 & cj \\
HD 178911B & HD 178911 B b    & 2552.0  & 0.340    & 0.124 & cj \\
HD 181296  & HR 7329 B        & 11124.0 & 0.000    & 0 & hj \\
HD 187085  & HD 187085 b      & 265.7   & 2.100    & 0.251 & cj \\
HD 189567  & HD 189567 b      & 8.5     & 0.111    & 0.189 & cl \\
           & HD 189567 c      & 7.0     & 0.197    & 0.160 & cl \\
HD 190228  & HD 190228 b      & 1726.8  & 2.405    & 0.559 & cj \\
HD 192263  & HD 192263 b      & 178.0   & 0.150    & 0.050 & cj \\
HD 196885  & HD 196885 A b    & 820.0   & 2.370    & 0.480 & cj \\
HD 197481  & AU Mic b         & 20.1    & 0.065    & 0.186 & hl \\
           & AU Mic c         & 9.6     & 0.110    & 0.041 & cl \\
           & AU Mic d         & 1.0     & 0.085    & 0 & hl \\
HD 202206  & HD 202206 (AB) c & 5689.2  & 2.410    & 0.220 & cj \\
HD 206860  & HN Peg b         & 6991.6  & 773.000  & 0 & cj \\
HD 206893  & HD 206893 c      & 4036.0  & 3.530    & 0.410 & cj \\
           & HD 206893 b      & 8899.0  & 9.600    & 0.140 & cj \\
HD 210277  & HD 210277 b      & 410.0   & 1.130    & 0.480 & cj \\
HD 215152  & HD 215152 b      & 1.8     & 0.058    & 0.357 & hl \\
           & HD 215152 c      & 1.7     & 0.067    & 0 & hl \\
           & HD 215152 d      & 2.8     & 0.088    & 0 & hl \\
           & HD 215152 e      & 2.9     & 0.154    & 0 & cl \\
HD 216435  & HD 216435 b      & 400.4   & 2.560    & 0.070 & cj \\
HD 218396  & HR 8799 e        & 3178.3  & 16.400   & 0.150 & cj \\
           & HR 8799 b        & 2000.0  & 68.000   & 0 & cj \\
           & HR 8799 c        & 3000.0  & 38.000   & 0.500 & cj \\
           & HR 8799 d        & 3000.0  & 24.000   & 0.600 & cj \\
BD-07 4003 & GJ 581 e         & 1.7     & 0.028    & 0.125 & hl \\
           & GJ 581 b         & 15.8    & 0.041    & 0 & hl \\
           & GJ 581 c         & 5.5     & 0.072    & 0 & hl \\
Ross 128   & Ross 128 b       & 1.4     & 0.050    & 0.116 & hl \\
%		\endfirsthead
%		\caption{continued.}\\
%		\hline
			Host Name   & Planet Name  & M$_{p}$  &  a  &   e  & Classification  \\
%		&                & (${M_ \oplus}$)     & (AU)  &    &   \\
		\hline
%		\endhead\\
%		\hline
%		\endfoot
%		\input{table3.dat}	
	\end{longtable}
    Notes. Columns 1 to 5 are the name of host stars, name, mass, orbit semi-major axis, and 
    eccentricity of planet, with data from NASA Exoplanet Archive. Column 6 is the planetary 
    classification: l = low-mass planet ; j = Jupiter; c = cool planet; h = hot planet.

\end{document}